# Revisiting Fermat's Factorization for the RSA Modulus

Sounak Gupta[1] and Goutam Paul[2]

[1] Department of Information Technology,
Bengal Engineering and Science University, Shibpur
Howrah 711 103, India.
`sounak.besu@gmail.com`
[2] Department of Computer Science and Engineering,
Jadavpur University, Kolkata 700 032, India.
`goutam.paul@ieee.org`

**Abstract.** We revisit Fermat's factorization method for a positive integer $n$ that is a product of two primes $p$ and $q$. Such an integer is used as the modulus for both encryption and decryption operations of an RSA cryptosystem. The security of RSA relies on the hardness of factoring this modulus. As a consequence of our analysis, two variants of Fermat's approach emerge. We also present a comparison between the two methods' effective regions. Though our study does not yield a new state-of-the-art algorithm for integer factorization, we believe that it reveals some interesting observations that are open for further analysis.

## 1 Introduction

Integer factorization is a classic problem in computational number theory. Given a positive integer $n$, factorization yields two positive integers $a > 1$, $b > 1$, such that $n = ab$. With the advancement of digital computers, there has been a considerable progress towards solving this problem in recent times. RSA cryptosystem [8] utilizes the concept of trapdoor functions for developing a public key encryption technique and is based on the fact that it is easy to multiply two large prime numbers but it is extremely difficult to obtain these primes by factoring their product.

Factoring algorithms are of two types: special purpose and general purpose algorithms. The efficiency of special purpose algorithms depends on the unknown factors, whereas the efficiency of the latter depends on the number to be factored. Some of the most important special purpose factoring algorithms are: *trial division*, *Pollard's rho method* [5], *Pollard's $p - 1$ method* [6], the *elliptic curve method* [1], *Fermat's method* [4], *squfof* [7] etc. *Quadratic sieve* [2] is a generic approach for developing general purpose algorithms. The most efficient general purpose algorithm known so far is the *number field sieve* [3]. Special purpose algorithms perform well for numbers with small factors, unlike the numbers used in the RSA. Therefore, general purpose factoring algorithms are the more important ones in the context of cryptographic systems and their security.

In this paper, we revisit Fermat's factorization from a new perspective. Fermat's method for factoring an odd integer $n$ consists of finding $n = x^2 - y^2$ where $x$ and $y$ are integers. One finds in succession $x = \lceil n^{0.5} \rceil, \lceil n^{0.5} \rceil + 1, \ldots$ and determines whether the difference $x^2 - n$ is a square or not. If $p$ and $q$ are primes and $n = pq$, then Fermat's method is quite efficient if $\frac{p}{q}$ is near 1, but it requires a large number of trials if $\frac{p}{q}$ is not near 1. However, in the later stages of this paper, we will prove that our approach provides better results than Fermat's method in certain regions even though we based our initial approach on the latter.

## 2 The Basic Method

Let $n = pq$ where $p$ and $q$ are large primes separated by at least a considerably large distance.

$$n = pq = \left(\frac{p+q}{2}\right)^2 - \left(\frac{p-q}{2}\right)^2. \tag{1}$$

Let $X_0 = \lceil \sqrt{n} \rceil$, $P_0 = X_0^2 - n$, $X_c = X_0 + c$, $P_c = X_c^2 - n$, where $c \geq 0$ is an integer. Note that $X_0$ and $P_0$ are both positive integers and fixed for a particular value of $n$. Since $X_c^2 - n = (X_0 + c)^2 - (X_0^2 - P_0) = c^2 + 2X_0 c + P_0$, we have

$$P_c = c^2 + 2X_0 c + P_0. \tag{2}$$

Now, if we compare $n = X_c^2 - P_c$ with Equation 1, we can say that we need to make $P_c$ a perfect square. Then, $P_c = (\frac{p-q}{2})^2$ and $X_c = \frac{p+q}{2}$. Hence, $p = X_0 + c + \sqrt{c^2 + 2X_0 c + P_0}$ and $q = X_0 + c - \sqrt{c^2 + 2X_0 c + P_0}$.

Thus, the smallest possible value of $P_c (= c^2 + 2X_0 c + P_0)$, which is a perfect square, gives us the ability to factorize large integers.

## 3 Further Analysis and a New Method

In Equation 2, the required value of $c$ is such that $P_c$ is a perfect square. So, it can be assumed that $P_c = (c + \alpha)^2$ where $\alpha \in \mathbf{N}$. From Equation 2, we have $(c + \alpha)^2 = c^2 + 2X_0 c + P_0$. Therefore,

$$c = \frac{\alpha^2 - P_0}{2(X_0 - \alpha)}. \tag{3}$$

Since, $c \geq 0$, $(\alpha^2 - P_0)$ and $(X_0 - \alpha)$ must both be positive or negative together.

1. When both are negative, then $\alpha^2 \leq P_0$ and $\alpha > X_0$, i.e., $0 \leq \alpha \leq \lfloor \sqrt{P_0} \rfloor$ and $\alpha > X_0$ (since, $\alpha \geq 0$). Since $\lfloor \sqrt{P_0} \rfloor \ll X_0$, the ranges are disjoint. So we have a contradiction. There is no possibility of $(\alpha^2 - P_0)$ and $(X_0 - \alpha)$ being negative together.

2. When both are positive, then $\alpha^2 > P_0$ and $\alpha < X_0$, i.e., $\lceil\sqrt{P_0}\rceil \leq \alpha < X_0$ (since, $\alpha \geq 0$). Hence, the range is found to be acceptable. So the range of $\alpha$ is $\lceil\sqrt{P_0}\rceil \leq \alpha < X_0$.

### 3.1 Nature of $\alpha$

From Equation 3, it is evident that $(\alpha^2 - P_0)$ must be even. So, $\alpha$ and $P_0$ should be both even or both odd. Also, since for RSA type values of $n$ ($n$ odd), $X_0$ and $P_0$ are opposite in nature (here nature refers to oddness or evenness), $\alpha$ and $X_0$ are opposite in nature and $(X_0 - \alpha)$ is always odd.

So, the number of test values of $\alpha$ is actually halved since, for a particular value of $n$, $\alpha$'s nature is fixed. We rewrite Equation 3 as
$c = \frac{\alpha^2 - P_0}{2(X_0-\alpha)} = \frac{\alpha^2 - X_0^2}{2(X_0-\alpha)} + \frac{X_0^2 - P_0}{2(X_0-\alpha)} = -\frac{X_0+\alpha}{2} + \frac{n}{2(X_0-\alpha)}$.

In the above equation, since $c \in \mathbf{N}$, then $(X_0 - \alpha)$ must divide $n$. Since, we already know $n = pq$, where $p$ and $q$ are prime numbers, $(X_0 - \alpha)$ is the smaller of the two prime numbers.

### 3.2 Test Case Reduction of $\alpha$

We know that the range of $\alpha$ is $\lceil\sqrt{P_0}\rceil \leq \alpha < X_0$ and as discussed in the previous section, the number of test cases in the range is halved. Thus, $\frac{X_0 - 1 - \lceil\sqrt{P_0}\rceil}{2}$ is the number of test cases. Also, we know that $(X_0 - \alpha)$ is the smaller of the numbers $p$ and $q$. So, judging by the value of $n$, we can predict the value of last digit of the two factors. Thus, we can actually guess the last digit of $\alpha$.

Now, from 0 to 9, there are five odd and five even numbers. Depending upon the nature of $\alpha$, five numbers qualify for being the last digit in any test case value of $n$. But $(X_0 - \alpha)$ can never have 5 as its last digit. Hence, we can say that number of possible last digits of $\alpha$ is 4. This is the maximum condition for number of valid values of $\alpha$'s last digit.

So, the number of test values of $\alpha$ is actually reduced to $\frac{4(X_0-1-\lceil\sqrt{P_0}\rceil)}{10} = \frac{2(X_0-1-\lceil\sqrt{P_0}\rceil)}{5}$.

### 3.3 Nature of $\alpha$-$c$ Curve

Equation 3 represents $c$ as a function of $\alpha$. In this section, we try to explore the nature of this function. This information is essential in the later sections of this paper.

We have $c = f(\alpha) = \frac{\alpha^2 - P_0}{2(X_0-\alpha)}$.

Differentiating $f(\alpha)$ w.r.t $\alpha$, we get

$$f'(\alpha) = \frac{-\alpha^2 + 2X_0\alpha - P_0}{2(X_0-\alpha)^2} \quad (4)$$

We can write $f'(\alpha) = \frac{\alpha}{2(X_0-\alpha)} + \frac{X_0\alpha - P_0}{2(X_0-\alpha)^2}$ and thereby it follows that $f'(\alpha)$ is always positive in the range $\lceil\sqrt{P_0}\rceil \leq \alpha < X_0$. Hence, $c = f(\alpha)$ is a monotonically increasing function in the derived interval.

## 3.4 An Alternative Approach

Here we present an alternative approach of analysis for the choices of $c$. Further analysis of this approach needs to be done to study its usefulness. We rewrite Equation 3 as

$$c = \frac{\alpha^2 - P_0}{2(X_0 - \alpha)} = \frac{X_0}{2} \cdot \frac{(\frac{\alpha}{X_0})^2 - \frac{P_0}{X_0^2}}{1 - \frac{\alpha}{X_0}}. \qquad (5)$$

Let $\frac{\alpha}{X_0} = \rho$ and $\frac{P_0}{X_0^2} = k$. Since, $\lceil \sqrt{P_0} \rceil \leq \alpha < X_0$, we have $\sqrt{k} < \rho < 1$. Also, $0 \leq P_0 \leq X_0^2 - (X_0 - 1)^2$, i.e., $0 \leq P_0 \leq 2X_0 - 1$. Therefore, $k = \frac{P_0}{X_0^2} \leq \frac{2X_0 - 1}{X_0^2} < \frac{2X_0}{X_0^2}$, i.e., $k < \frac{2}{X_0}$. Hence, $0 < k < \frac{2}{X_0}$. For a particular value of $n$, $k$ is always constant. $k$ is typically very close to 0.

Replacing these variables in Equation 5, we get $c = f_k(\rho) = \frac{X_0}{2} \cdot \frac{\rho^2 - k}{1 - \rho} = \frac{X_0}{2} \left( \frac{1-k}{1-\rho} - (1 + \rho) \right)$.

Since, we already know that $c$ is a positive integer, there might be some scope here for predicting the approximate value of $\rho$ for which $c$ becomes an integer. Further mathematical exploration is essential for extracting any meaningful results from this approach.

## 3.5 Practical Range Determination for $\alpha$

In the beginning of Section 3, we deduced the range of $\alpha$ for which large integer factorization is possible. So far, primality of the factors have not been considered. In this section, we further explore the concept in the context of RSA cryptosystem. We assume $n = pq$, $q < p < 2q$, where $p$ and $q$ are two prime numbers. In Section 2, we derived that

$p = X_c + \sqrt{P_c}$ and $q = X_c - \sqrt{P_c}$, where $P_c = c^2 + 2X_0 c + P_0$.

Using these expressions in the relation $q < p < 2q$, we get

$X_c - \sqrt{P_c} < X_c + \sqrt{P_c} < 2(X_c - \sqrt{P_c})$.

That is, $-\sqrt{P_c} < \sqrt{P_c} < X_c - 2\sqrt{P_c}$.

Now, $P_c \geq P_0$ as $c \geq 0$, and hence we can say $\sqrt{P_c} \geq \sqrt{P_0}$.

So, the relation becomes $\sqrt{P_0} \leq \sqrt{P_c} < X_c - 2\sqrt{P_c}$.

Again, $\sqrt{P_c} < X_c - 2\sqrt{P_c} \Leftrightarrow 3\sqrt{P_c} < X_c \Leftrightarrow 9P_c < X_c^2$ (squaring both sides)
$\Leftrightarrow 9(c^2 + 2X_0 c + P_0) < (X_0 + c)^2 \Leftrightarrow c^2 + 2X_0 c + P_0 - \frac{n}{8} < 0$ (Since, $X_0^2 - P_0 = n$).

Boundary values of $c = -X_0 \pm \sqrt{n + \frac{n}{8}}$. Since, $c \geq 0$, we have

$0 \leq c < -X_0 + \sqrt{\frac{9n}{8}} \Leftrightarrow 0 \leq c < -X_0 + 1.061 X_0$ (considering $\sqrt{n}$ as $X_0$).

Hence, $0 \leq c < 0.061 X_0$. The range is approximate in calculation.

From Equation 3, we get

$\frac{\alpha^2 - P_0}{2(X_0 - \alpha)} < 0.061 X_0$

$\Leftrightarrow \alpha^2 + 0.122 X_0 \alpha - (P_0 + 0.122 X_0^2) < 0$.

Boundary values of $\alpha = -0.061 X_0 \pm X_0 \cdot \sqrt{0.125721 + k}$ (as assumed in Section 3.4, $\frac{P_0}{X_0^2} = k$ which is constant for a particular value of $n$). $\alpha$ has already been proved to be $\geq \lceil\sqrt{P_0}\rceil$. Hence, $\lceil\sqrt{P_0}\rceil \leq \alpha < X_0 \cdot (\sqrt{0.125721 + k} - 0.061)$. We know that $k \leq \frac{2}{X_0}$ and so its value is very small. Thus, it can be considered that $\sqrt{0.125721 + k} < 0.36$

Hence,
$$\lceil\sqrt{P_0}\rceil \leq \alpha < 0.3X_0. \tag{6}$$

The range is approximate in calculation.

## 4 Comparative Study of the Two Methods' Effective Regions

So far we have discussed two different approaches. In the first one, we try to calculate the value of $c$ through several iterations. Let us designate this method as *c-method*. In the second one, we try to calculate the value of $\alpha$ through several iterations. Let us call this method the *$\alpha$-method*. The objective here is to determine the region of effectiveness of the two methods. Let us first summarize the steps involved in both the methods.

1. *c*-method: $c^2, 2X_0c, c^2 + 2X_0c, c^2 + 2X_0c + P_0, x = \sqrt{c^2 + 2X_0c + P_0}, y = x - \lfloor x \rfloor$, Increment $c$ until $y = 0$.
2. $\alpha$-method: $\alpha^2, \alpha^2 - P_0, 2\alpha, 2X_0 - 2\alpha, x = \frac{\alpha^2 - P_0}{2(X_0 - \alpha)}, y = x - \lfloor x \rfloor$, Increment $\alpha$ until $y = 0$.

Thus, in terms of number of steps, both methods are equally effective. In general, comparative effectiveness can be determined by calculating the minimum number of consecutive iterations of $c$ needed to match the progress of a certain number of iterations of $\alpha$. We know from Section 3.2 that out of every 10 consecutive integer values of $\alpha$, we need to consider only four of them. Since, in the *c*-method, we increment $c$ by 1, we can say that $\frac{dc}{d\alpha} \geq 0.4$. Hence, from Equation 4, we get

$\frac{-\alpha^2 + 2X_0\alpha - P_0}{2(X_0 - \alpha)^2} \geq 0.4$

$\Leftrightarrow 9\alpha^2 - 18X_0\alpha + 9P_0 + 4(X_0^2 - P_0) \leq 0$

$\Leftrightarrow \alpha^2 - 2X_0\alpha + (P_0 + \frac{4n}{9}) \leq 0 (since, X_0^2 - P_0 = n)$.

Boundary values of $\alpha = X_0 \pm \sqrt{(X_0^2 - P_0) - \frac{4n}{9}} = X_0 \pm \sqrt{\frac{5n}{9}}$

$\approx X_0 \pm X_0\sqrt{\frac{5}{9}}$ (considering $\sqrt{n}$ as $X_0$) $\approx X_0(1 \pm 0.745)$.

i.e., $0.255X_0 \leq \alpha \leq 1.745X_0$.

Since, we know $\lceil\sqrt{P_0}\rceil \leq \alpha < X_0$, hence, $\alpha$-method is more effective in the region $0.255X_0 \leq \alpha < X_0$ and *c*-method is more effective in the region $\lceil\sqrt{P_0}\rceil \leq \alpha \leq 0.255X_0$.

From Equation 6 we can say that for RSA-specific purposes, $\alpha$-method is more effective in the region $0.255X_0 \leq \alpha < 0.3X_0$ and the *c*-method is effective in the region $\lceil\sqrt{P_0}\rceil \leq \alpha \leq 0.255X_0$. The ranges are approximate in calculation.

## 5 An Illustrative Example

Let us consider a 264-bit integer
$n = 24758167959654528007156374531915464081839760935532218683970$
$8649238085888673119$ with factors
$p = 6847944682037444681162770672798288913849$ and
$q = 3615415881585117908550243505309785526231$.
We have $X_0 = 49757580286479494366940039692986641174 73$,
$P_0 = 31716812982186337037801065018400552326 10$,
$c = 25592225316333185816250311975537310256 7$,
$\alpha = 13603421470628315281437604639888785912 42$.

The lower bound of the interval in which the $\alpha$-method gives better results is
$0.255 X_0 = 126881829730522710635697101217115934 9956$.

Number of iterations for the $\alpha$-method is
$z = \alpha - 0.255 X_0 = 9152384975760442178678945181771924 1286$.

Considering sieving, the number of iterations is
$0.4z = 36609539903041768714715780727087696 514$.

Thus, the difference between the number of test cases for the $c$-method and the $\alpha$-method is $c - 0.4z = 219312713260290089447787339028285406053$.

## References

1. H. W. Lenstra, Jr. Factoring integers with elliptic curves. *Annals of Mathematics*, pages 649-673, vol. 126, 1987.
2. Integer Factoring. A. K. Lenstra. *Designs, Codes and Cryptography*, pages 101-128, vol. 19, no. 2/3, 2000.
3. A. K. Lenstra and H. W. Lenstra. The Development of the Number Field Sieve. Lecture Notes in Mathematics, vol. 1554, Springer-Verlag, 1993.
4. James McKee. Speeding Fermat's Factoring Method. Mathematics of Computation, pages 1729-1737, vol. 68, no. 228, March 1, 1999.
5. J. M. Pollard. A Monte Carlo method for factorization. *BIT*, pages 331-334, vol. 15, 1975.
6. J. M. Pollard. Theorems on factorization and primality testing. Proc. Cambridge Philos. Soc., pages 521-528, vol. 76, 1974.
7. R. Schoof. Quadratic fields and factorization. Computational methods in number theory, Math. Centre Tracts, Mathematisch Centrum, Amsterdam, 1983, pages 235-286, vol. 154/155.
8. R. L. Rivest, A. Shamir, and L. Adleman. A method for obtaining digital signatures and public key cryptosystems. *Communications of the ACM*, pages 120-126, vol 21, no. 2, February, 1978.